\documentclass[11pt,a4paper]{article}
\usepackage{jcappub}

\usepackage{amssymb}
\usepackage{amsmath}
\usepackage{graphicx}

\title{Cosmological Model-independent Gamma-ray Bursts Calibration and its Cosmological Constraint to Dark Energy}

\author{Lixin Xu}
\emailAdd{lxxu@dlut.edu.cn}

\affiliation{Institute of Theoretical Physics, School of Physics \&
Optoelectronic Technology, Dalian University of Technology, Dalian,
116024, P. R. China}

\affiliation{College of Advanced Science \& Technology, 
Dalian University of Technology, Dalian, 116024, P. R. China}

\affiliation{Korea Astronomy and Space Science Institute,
Yuseong Daedeokdaero 776,
Daejeon 305-348,
R. Korea}

\abstract{
As so far, the redshift of Gamma-ray bursts (GRBs) can extend
to $z\sim 8$ which makes it as a complementary probe of dark energy
to supernova Ia (SN Ia). However, the calibration of GRBs is still a big
challenge when they are used to constrain cosmological models. Though, the absolute magnitude of
GRBs is still unknown, the slopes of GRBs correlations can be used as a useful constraint to dark energy in a completely cosmological model independent way. In
this paper, we follow Wang's model-independent distance measurement
method and calculate their values by using $109$ GRBs events via the
so-called Amati relation. Then, we use the obtained model-independent distances to constrain $\Lambda$CDM model as an example.}

\arxivnumber{1005.5055}

\begin{document}

\maketitle

\section{Introduction}

Since the discovery of an accelerated
expansion of our Universe through the observations of supernova Ia (SN Ia) \cite{ref:Riess98,ref:Perlmuter99}, many
cosmic observations have been used to explore the nature of dark energy which has negative pressure and pushes the Universe into an accelerated phase. Particularly, the redshift of Gamma-ray burst (GRBs) can extend to higher redshift $z\sim 8$. This merit makes it as a complementary cosmic probe to SN Ia. However, a big challenge, the so-called circular problem, has to be overcome when one calibrates the GRBs to obtain the distances at different redshifts. In \cite{ref:Schaefer}, Schaefer {\it et. al.} have derived the distance modulus in $\Lambda$CDM model by using five GRBs correlations. The so-called circular problem would be committed  when one uses the resulted distance redshift relation to constrain any other cosmological models beyond $\Lambda$CDM. To overcome this circular problem, Li, {\it et. al} \cite{ref:Lihong} put the GRBs correlation and cosmological model constraint together. Via the Markov Chain Monte Carlo (MCMC) method, they fixed the calibration parameters and constrained the model parameter space simultaneously. However, it looks like using a 'ruler' having no marks to measure the length of an object. And after the measurement, the length and scale of ruler are given together. So the lack of calibration makes the GRBs lose the power to constrain cosmological models. Cosmography method  was considered in \cite{ref:cosmography} by parameterizing the luminosity distance $d_L$ in terms of deceleration $q_0$, jerk $j_0$ and snap $s_0$ parameters. In this way, the cosmological model dependent problem was removed. Liang {\it
et. al.} \cite{ref:Liang} calibrated GRBs by using low redshift SN Ia and obtained a tight constrain to the cosmological model parameter space. This method was reconsidered by Wei \cite{ref:Wei,ref:Wei109}. By analyzing the calibration process carefully, one can find the potential drawback. The calibrated GRBs correlation makes GRBs have the same luminosity distance-redshifts relation as SN Ia at lower redshifts. In this way, the relation is extended to higher redshifts. Equivalently, one just extends the luminosity distance-redshifts relation of SN Ia to higher redshifts. So, it makes the obtained luminosity distance-redshifts relation strongly depends on that of SN Ia. The worst thing is that it makes the data points of GRBs useless, because one has known the luminosity distance-redshifts relation at high redshifts from SN Ia, though no SN Ia is found at the high redshift regions of GRBs. Alternatively, Wang presented a model-independent distance measurement from GRBs calibrated internally \cite{ref:wang}. The main point of Wang's method is that the statistical errors of correlation parameters $\sigma_a$, $\sigma_b$
and systematic error $\sigma_{sys}$ obtained in $\Lambda$CDM models are used, but are not the correlation parameters $a$ and $b$ themselves, for the definitions of $a$, $b$ please see Eq. (\ref{eq:calib}). The viability of this implement comes from the observations that the errors of correlation parameters are almost the same for different values of $\Omega_{m0}$ for $\Lambda$CDM model, though the values of $a$ and $b$ are really different. Then, in terms of a set of model-independent distance measurements, the cosmic constraint from GRBs is set up via
cubic spline interpolation from cosmological model independent
distance ratio $\bar{r}_p(z_i)$. The merits of this method are
follows: (i) the constraint from GRBs is in a cosmological model
independent way. It alleviates the circular problem. So, it can be used to constrain any other cosmological
models. (ii) It is not calibrated by any other external data sets.
It does not suffer any consistent problem when it is combined with
other data sets as cosmic constraints. (iii) The cosmological model independent calibration is done firstly. It means that the 'ruler' has been marked. (iv) Though the absolute magnitude of GRBs is unkown, the slopes of GRBs correlations can be used as cosmological constraints. 

Recently, Wei \cite{ref:Wei109} used $109$ GRBs data points via
Amati relation \cite{r16,r17,r18} calibrated by SN Ia to
constrain cosmological model. Based on the points mentioned above, to alleviate the data sets dependence and 
circular problem, we shall present our calculation results based on Wang's method via Amati correlation. In fact, we find a new $\chi^2_{GRB}$ which only depends on the slopes of GRBs correlation and make the absolute magnitude of GRBs irrelevant. For the details, please see section \ref{sec:method}. As results, a set of model-independent distance measurements are obtained on the basis of Amati relation that can be used to constrain cosmological models. 

However we have to stress that the Amati relation has been criticized for many reasons in the literatures. Li demonstrated an ambiguity in determining the redshifts of GRBs \cite{ref:Li}. Of course, this ambiguity can be overcome when the redshift is well determined. The major criticism came from Nakar and Prian \cite{ref:Nakar} who developed a test for the Amati relation even in the case where the redshifts of GRBs were unknown. The test was also generalized by Band and Preece \cite{ref:Band}. They concluded that the Amati relation suffered the problem of selection bias. Recently, the authors of \cite{ref:Collazzi} have also concluded that the Amati relation is an artifact of selection effects with the burst population and the detector. They also point out that the Amati relation is failed whether or not the bursts have measured spectroscopic redshifts. If this is true for Amati relation, the results obtained based on Amati relation would be unreliable. But it is still in debating. So, before the dust settles down, we still assume that Amati relation is reliable in this paper.

This paper is structured as follows. In section \ref{sec:method}, the values of
cosmological parameter dependance in $\Lambda$CDM model are presented
as shown in \cite{ref:wang}. The errors of correlation parameters
$\sigma_a$, $\sigma_b$ and systematic error $\sigma_{sys}$ are also
calculated via Amati relation with $109$ GRBs data points. The $5$ bins
model-independent distance measurements will also be found in this section. In section III, we use the resulted data points  to constrain $\Lambda$CDM model as an example. A summary and
discussion are put in section IV.

\section{Calibration of GRBs and Model-independent Distance Measurement} \label{sec:method}

Following the work of \cite{ref:Schaefer}, we consider the well-known Amati
$E_{p,i}-E_{iso}$ correlation \cite{ref:Amati'srelation,r16,r17,r18}
in GRBs, where $E_{p,i}=E_{p,obs}(1+z)$ is the cosmological
rest-frame spectral peak energy, and $E_{iso}$ is the isotropic
energy
\begin{equation}
E_{iso}=4\pi d^2_LS_{bolo}/(1+z)
\end{equation}
in which $d_L$ and $S_{bolo}$ are the luminosity distance and the
bolometric fluence of the GRBs respectively. Following
\cite{ref:Schaefer}, we rewrite the Amati relation in the form of
\begin{equation}
\log\frac{E_{iso}}{{\rm erg}}=a+b\log\frac{E_{p,i}}{300{\rm
keV}}.\label{eq:calib}
\end{equation}

One fitts the Amati relation through the minimization of $\chi^2$ which is given
by \cite{ref:Schaefer}
\begin{equation}
\chi^2=\sum^N_{i=1}\frac{y_i-a-bx_i}{\sigma^2_{y,i}+b^2\sigma^2_{x,i}},
\end{equation}
where
\begin{eqnarray}
x_i&=&\log\frac{E_{p,i}}{300{\rm keV}}\\
y_i&=&\log\frac{E_{iso}}{{\rm erg}}=\log\frac{4\pi
S_{bolo,i}}{1+z}+2\log\bar{d}_L
\end{eqnarray}
where $\bar{d}_L$ is defined as \cite{ref:wang}
\begin{equation}
\bar{d}_L=H_0(1+z)r(z)/c,
\end{equation}
and the errors are calculated by
using the error propagation law \cite{ref:errors}:
\begin{eqnarray}
\sigma_{x,i}&=&\frac{\sigma_{E_{p,i}}}{\ln10E_{p,i}}\\
\sigma_{y,i}&=&\frac{\sigma_{S_{bolo,i}}}{\ln10S_{bolo,i}}.
\end{eqnarray}

By calculating the value of $\chi^2$, we find it is large and dominated by the systematic errors, and on the contrast 
the statistical errors on $a$ and $b$ are small. Following
\cite{ref:Schaefer}, the systematic error $\sigma_{sys}$ can be
derived by required $\chi^2=\nu$ (the degrees of freedom). At last,
the total error $\sigma^2_{tot}=\sigma^2_{stat}+\sigma^2_{sys}$ is obtained.
It would be noticed that in our case, the best fit value of $a$ will
be less $2\log(c/H_0)$ than that in the definition of luminosity distance
$d_L=(1+z)r(z)$ \cite{ref:wang}. For this definition, the value of
$H_0$ is absorbed into the calibration of GRBs because of the lack
of fixing the absolute magnitude of GRBs. Then, in our treatment the
results will be $H_0$ free.

As shown in \cite{ref:wang}, the calibration of GRBs is cosmological model dependent, because the values of $a$ and $b$ are obtained on the basis of $\Lambda$CDM model. So the calibrated result can not be used to constrain any other cosmological models. This the so-called well-known model dependent problem or circular problem. In Tab. \ref{Tab:calibration}, the cosmological parameter dependence are shown. With a careful observation, one can find that the $1\sigma$ errors bars of $a$, $b$ and system are almost the same which do not depend on the cosmological parameters. In this paper as suggested by Wang \cite{ref:wang}, we shall only use the
systematic error and the errors of $a$ and
$b$ in the case of $\Omega_{m0}=0.27$ for $\Lambda$CDM model as the
standard values. That is to say the values of $\sigma_a$, $\sigma_b$
and $\sigma_{sys}$ will be used in the following sections {\it not}
the values of $a$ and $b$ derived from $\Lambda$CDM model.

\begin{table}[htbp]
\begin{center}
\begin{tabular}{c|c|c|c}
\hline\hline
 &\ $\Omega_{m0}=0.27$ & $\Omega_{m0}=0.2$ & $\Omega_{m0}=0.4$ \\ \hline
$a$ & $-3.392\pm0.0368$ & $-3.343\pm0.0358$ & $-3.467\pm0.0364$ \\
$b$ & $1.583\pm0.0729$ & $1.600\pm0.0744$ & $1.554\pm0.0725$\\
$\sigma_{sys}$ & $0.324$ & $0.328$ & $0.321$\\
\hline\hline
\end{tabular}
\end{center}
\caption{\label{Tab:calibration} Systematic error and values of $a$
and $b$ for GRBs Amati relation in the cases of
$\Omega_{m0}=0.27,0.2,0.4$ in $\Lambda$CDM model.}
\end{table}

Following the work of Wang \cite{ref:wang}, the $\chi_{GRB}^2$ of a cosmological
model is given by
\begin{equation}
\chi_{GRB}^2=\sum^{N_{GRB}}_{i=1}\frac{[(\log\bar{d}^2_L)^{data}_i-\log\bar{d}^2_L(z_i)]^2}{[\sigma(\log\bar{d}^2_L)^{data}_i]^2},\label{eq:chi2GRBwang}
\end{equation}
where
\begin{eqnarray}
[\sigma(\log\bar{d}^2_L)^{data}_i]^2&=&\sigma_a^2+\left(\sigma_b\log\frac{E_{p,i}}{300{\rm
keV}}\right)^2
+\left(b\frac{\sigma_{E_{p,i}}}{\ln10E_{p,i}}\right)^2
+\left(\frac{\sigma_{S_{bolo,i}}}{\ln10S_{bolo,i}}\right)^2
+\sigma^2_{sys},\\
(\log\bar{d}^2_L)^{data}_i&=& a +
b\log\frac{E_{p,i}}{300{\rm keV}}-\log\frac{4\pi S_{bolo,i}}{1+z}.\\
\end{eqnarray}

To constrain a cosmological model, one uses a set of
model-independent distance measurements $\{\bar{r}_p(z_i)\}$:
\begin{equation}
\bar{r}_p(z_i)\equiv\frac{r_p(z)}{r_p(z_{0})},\quad r_p(z)\equiv
\frac{(1+z)^{1/2}}{z}\frac{H_0}{c}r(z),\label{eq:rp}
\end{equation}
where $r(z)=d_L(z)/(1+z)$ is the comoving distance at redshift $z$,
$z_{0}$ is the lowest GRBs redshift\footnote{$z_{0}=0.17$ was used
in \cite{ref:wang}. In this work, the lowest redshift of GRBs is
$z_0=0.0331$}. Here, the definition of $r_p$ is different from
Wang's definition \cite{ref:wang} $r_p(z)\equiv (1+z)^{1/2}H_0 r(z)/(zch)$
where $h=H_0/(100{\rm km s}^{-1}{\rm Mpc}^{-1})$. In our definition,
the distance measurement $\bar{r}_p(z_i)$ and $r_p(z)$
is completely $H_0$ free. It can be seen from the facts that the
definition of $r(z)$ is $r(z)=c/H_0\int^z_0 dz'/E(z')$ where
$H^2(z)=H^2_0 E^2(z)$ is the Hubble parameter. In terms of our definition,
$\bar{d}_L$ can be rewritten as
\begin{equation}
\bar{d}_L=z(1+z)^{1/2}r_p(z_{0})\bar{r}_p(z).
\end{equation}
We divide the redshifts of GRBs into $N$ bins, i.e.
$\{z_i\},i=1,...,N$, and assume the corresponding values of
$\{\bar{r}_p(z_i)\},i=1,...,N$ which {\it do not} depend on any
cosmological models. Then the values of $\bar{r}_p(z)$ at arbitrary
redshift $z$ can be obtained by cubic spline interpolation from
$\{\bar{r}_p(z_i)\}$. So, the values of $d_L(z)$ and $\bar{d}_L(z)$,
etc at redshift $z$ can be found easily. Given each set of
$\{\bar{r}_p(z_i)\},i=1,...,N$, we calibrate the GRBs and calculate
the likelihood simultaneously via Eq. (\ref{eq:chi2GRBwang}) by using Markov Chain Monte
Carlo (MCMC) method \cite{ref:MCMC}. The MCMC is a global fitting method which is used to determine the cosmological parameters. In adopting the MCMC approach, we generate using Monte Carlo method a chain of
sample points distributed in the parameter space according to the
posterior probability, using the Metropolis-Hastings algorithm with
uniform prior probability distribution. In the parameter space
formed by the constraint cosmological parameters, a random set of
initial values of the model parameters is chosen to calculate the
$\chi^2$ or the likelihood. Whether the set of parameters can be
accepted as an effective Markov chain or not is determined by the
Metropolis-Hastings algorithm. The accepted set not only forms a
Markov chain, but also provides a starting point for the next
process. We then repeat this process until the established
convergence accuracy can be satisfied.  The convergence is tested by
checking the so-called worst e-values [the
variance(mean)/mean(variance) of 1/2 chains] $R-1 < 0.005$ \cite{ref:MCMC}. As results, we obtain a set of distances
$\{\bar{r}_p(z_i)\},i=1,...,N$ which are {\it independent} on any
assumption of cosmological parameters. It comes from the
observations that, in the process of calibration, the
statistical and systematic errors of $\sigma_a$, $\sigma_b$ and
$\sigma_{sys}$ are only used. We {\it do not} use any values of $a$ and $b$ calibrated in $\Lambda$CDM model. The important thing is that
the statistical and systematic errors $\sigma_a$, $\sigma_b$ and
$\sigma_{sys}$ are almost model parameter independent. Thanks to this
feature, this method is model-independent. Here, in the MCMC
analysis, we take $a$, $b$ and $N$ $\{\bar{r}_p(z_i)\},i=1,...,N$ as
free parameters. So, the degree of freedoms is $\nu=109-2-N$. Once
these values of $\{\bar{r}_p(z_i)\},i=1,...,N$ are obtained, a
cosmological model can be constrained by GRBs via the $\chi^2$
\begin{eqnarray}
\chi^2_{GRB}&=&[\Delta\bar{r}_p(z_i)]\cdot(Cov^{-1}_{GRB})_{ij}\cdot[\Delta\bar{r}_p(z_i)],\label{eq:chi2GRB}\\
\Delta\bar{r}_p(z_i)&=&\bar{r}^{data}_p(z_i)-\bar{r}_p(z_i),
\end{eqnarray}
where $\bar{r}_p(z_i)$ is defined by Eq. (\ref{eq:rp}) and
$(Cov^{-1}_{GRB})_{ij},i,j=1...N$ is the covariance matrix. In this way, the constraints from larger observational GRBs data points are projected into relative smaller number of points.
Of course, this method can be generalized to discuss other problems.

Now, we present some discussion about the treatment of $r_p(z_0)$.
Here, we do not calculate the value of $r_p$ at the redshift $z_0$
via its definition (\ref{eq:rp}) according to any cosmological
model. Because, if we calculate the value for any cosmological model, a model or cosmological parameter dependence will be introduced again. The values of $r_p(z_0)$ are
fixed by the calibration relation (\ref{eq:calib}), i.e. via the
relation
\begin{equation}
r_p(z_0)=10^{a/2}(\frac{E_{p,i}(z_0)}{300})^{b/2}\frac{1}{z_0(4\pi
S_{bolo}(z_0))^{1/2}}.
\end{equation}
In fact, once this relation is used, one can recast  Eq.
(\ref{eq:chi2GRBwang}) into the following form
\begin{equation}
\chi_{GRB}^2=\sum^{N_{GRB}}_{i=1}\frac{(Y_i-Y_{0})^2}{(\sigma^{data}_{Y,i})^2},\label{eq:chi2newform}
\end{equation}
where $Y_i$ is defined as
\begin{equation}
Y_i=b\log E_{p,i}(z_i)-\log S_{bolo}(z_i)-2\log z_i
-2\log\bar{r}_p(z_i)
\end{equation}
and $Y_{0}=Y_i(z_0)$, the $\sigma^{data}_{Y,i}$ is the total
$1\sigma$ errors of data sets
\begin{equation}
\sigma^{data}_{tot,i}=\left(\sigma_b\log\frac{E_{p,i}}{300{\rm
keV}}\right)^2
+\left(b\frac{\sigma_{E_{p,i}}}{\ln10E_{p,i}}\right)^2
+\left(\frac{\sigma_{S_{bolo,i}}}{\ln10S_{bolo,i}}\right)^2
+\sigma^2_{sys}.
\end{equation}
One can see that the parameter $a$ is removed from this new form of
$\chi_{GRB}^2$. In other words, the information about the slope
$b$ of GRBs correlation is used alone. So the absolute magnitude $a$ of GRBs is
irrelevant in this method. Of cause, the $r_p(z_0)$ can be fixed
by consulting a special cosmological model or other data sets. But in
that way, the circular problem and data sets relevance problem will
come back. Equivalently, the absolute magnitude of GRBs was fixed in a special cosmological model. So, for every possible values of $b$ in every running of MCMC, the $\chi_{GRB}^2$ is
calculated. Here, we must keep in mind that the values of
parameter $b$ are not taken from Tab. \ref{Tab:calibration}. It is treated as a free parameter. So, this
method is completely cosmological parameter in-dependent and self consistent.

\begin{figure}[tbh]
\centering
\includegraphics[width=5.0in]{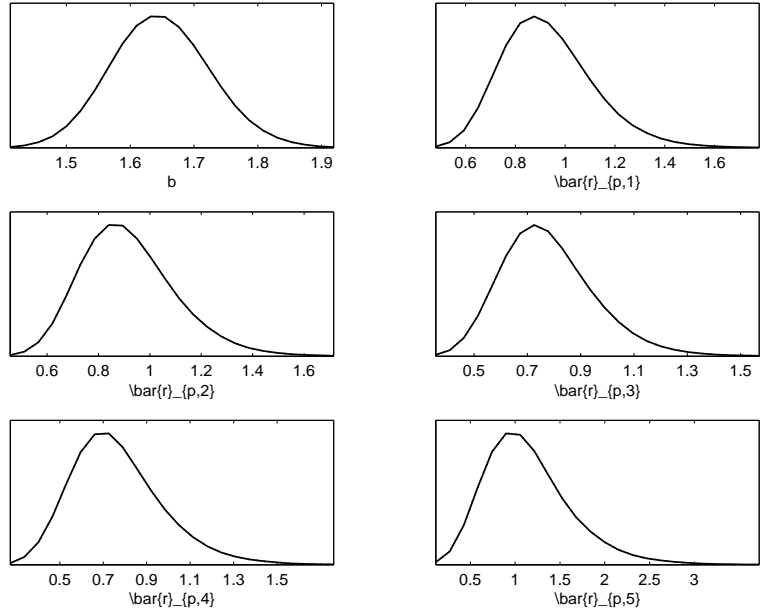}
\caption{The $1D$ marginalized probabilities of model
parameters.}\label{fig:brps}
\end{figure}

\begin{figure}[tbh]
\centering
\includegraphics[width=5.0in]{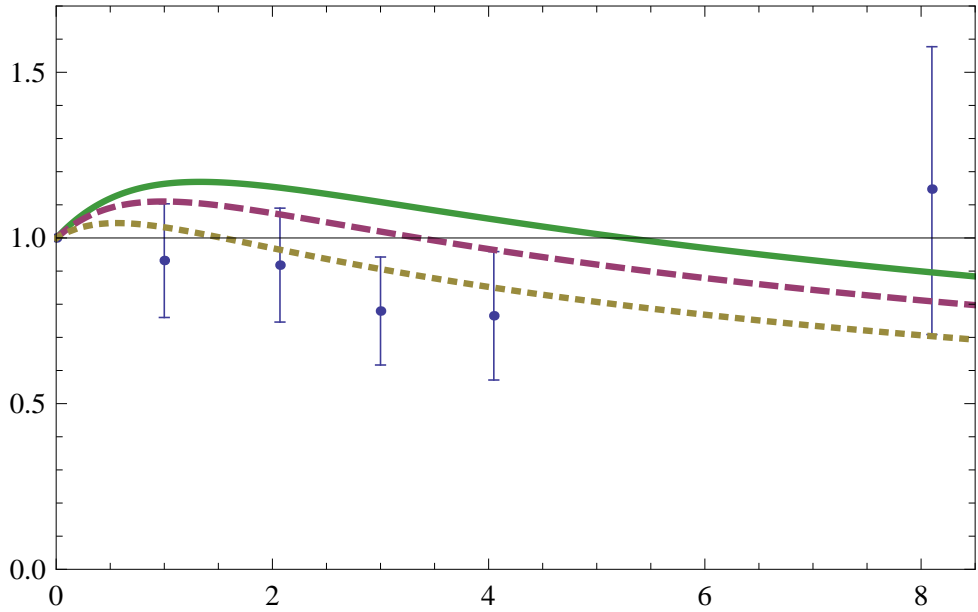}
\caption{The mean values of distance measurement from $109$ GRBs via
Amati relation with $1\sigma$ error bars. The lines from up to
bottom correspond to $\Omega_{m0}=0.2,0.27,0.4$ respectively.
}\label{fig:errorbars}
\end{figure}

We divide the redshifts into $N=5$ bins and run the MCMC codes which is based on the publicly available {\bf CosmoMC} package \cite{ref:MCMC}]. The
chains have worst e-value $R-1 = 0.0017$ which is much smaller than $0.005$. The
resulted model-independent distances and covariance matrix from
$109$ GRBs are shown in Tab. \ref{tab:distance}
\begin{table}[htbp]
\begin{center}
\begin{tabular}{c|c|c|c|c}
\hline\hline
 & $z$ & $\bar{r}^{data}_p(z)$ & $\sigma(\bar{r}_p(z))^+$ &  $\sigma(\bar{r}_p(z))^-$\\ \hline
$0$ & $\quad0.0331\quad$ & $\quad1.0000\quad$ & $-$ & $-$  \\
$1$ & $1.0000$  & $0.9320$ & $0.1711$ & $0.1720$ \\
$2$ & $2.0700$  & $0.9180$ & $0.1720$ & $0.1718$ \\
$3$ & $3.0000$  & $0.7795$ & $0.1630$ & $0.1629$ \\
$4$ & $4.0480$  & $0.7652$ & $0.1936$ & $0.1939$ \\
$5$ & $8.1000$  & $1.1475$ & $0.4297$ & $0.4389$ \\
  \hline\hline
\end{tabular}
\end{center}
\caption{\label{Tab:datapoints} Distances measured form $109$ GRBs
via Amati relation with $1\sigma$ upper and lower
uncertainties. }\label{tab:distance}
\end{table}
and Eq. (\ref{eq:covM}). As already mentioned above, $z_{0}=0.0331$
is adopted in this work. The $\{\bar{r}_p(z_i)\},i=1,...,5$
correlation matrix is given by
\begin{eqnarray}
&&(\overline{Cov}_{GRB})= \left(
\begin{array}{ccccc}
$1.0000$ & $0.7780$ & $0.8095$ & $0.6777$ & $0.4661$ \\
$0.7780$ & $1.0000$ & $0.7260$ & $0.6712$ & $0.3880$ \\
$0.8095$ & $0.7260$ & $1.0000$ & $0.6046$ & $0.5032$ \\
$0.6777$ & $0.6712$ & $0.6046$ & $1.0000$ & $0.1557$ \\
$0.4661$ & $0.3880$ & $0.5032$ & $0.1557$ & $1.0000$
\end{array}
\right),
\end{eqnarray}
and the corresponding covariance matrix is given by
\begin{equation}
(Cov_{GRB})_{ij}=\sigma(\bar{r}_p(z_i))\sigma(\bar{r}_p(z_j))(\overline{Cov}_{GRB})_{ij},\label{eq:covM}
\end{equation}
where
\begin{eqnarray}
\sigma(\bar{r}_p(z_i))=\sigma(\bar{r}_p(z_i))^+, \quad {\rm if}\quad
\bar{r}_p(z)\geq
\bar{r}_p(z)^{data}; \\
\sigma(\bar{r}_p(z_i))=\sigma(\bar{r}_p(z_i))^-, \quad {\rm if}\quad
\bar{r}_p(z)< \bar{r}_p(z)^{data},
\end{eqnarray}
here the $\sigma(\bar{r}_p(z_i))^+$ and $\sigma(\bar{r}_p(z_i))^-$ are
the $1\sigma$ errors listed in Tab. \ref{Tab:datapoints}. The
marginalized values and the corresponding upper and lower bounds are
used when GRBs is used as a cosmological constraint.  The $1D$
distributions of model parameters are shown in Fig.\ref{fig:brps}.
The distance measurements from $109$ GRBs via Amati correlation with
$1\sigma$ error bars are shown in Fig. \ref{fig:errorbars}, where
the solid lines correspond to $\Lambda$CDM model with different
values of $\Omega_{m0}=0.2,0.27,0.4$ from up to bottom respectively.

The data points shown in Tab. \ref{tab:distance} and Fig. \ref{fig:errorbars} are totally cosmological model independent, so one can use them to constrain any other cosmological models. In the next section, as an example, these obtained data points will be used to constrain $\Lambda$CDM model.  From Fig. \ref{fig:errorbars}, one can find that GRBs favors large values of $\Omega_{m0}$ for $\Lambda$CDM model. This point can be confirmed in the next section. And the concordance model ($\Omega_{m0}=0.27$) is almost at the boundary of $1\sigma$ regions of data points. So the null hypothesis that Amati relation is based on cosmology can be rejected slightly greater than $4.5\sigma$ as shown in Fig. \ref{fig:errorbars}.

\section{Cosmological Constraint to Dark Energy}

\begin{figure}[tbh]
\centering
\includegraphics[width=5.0in]{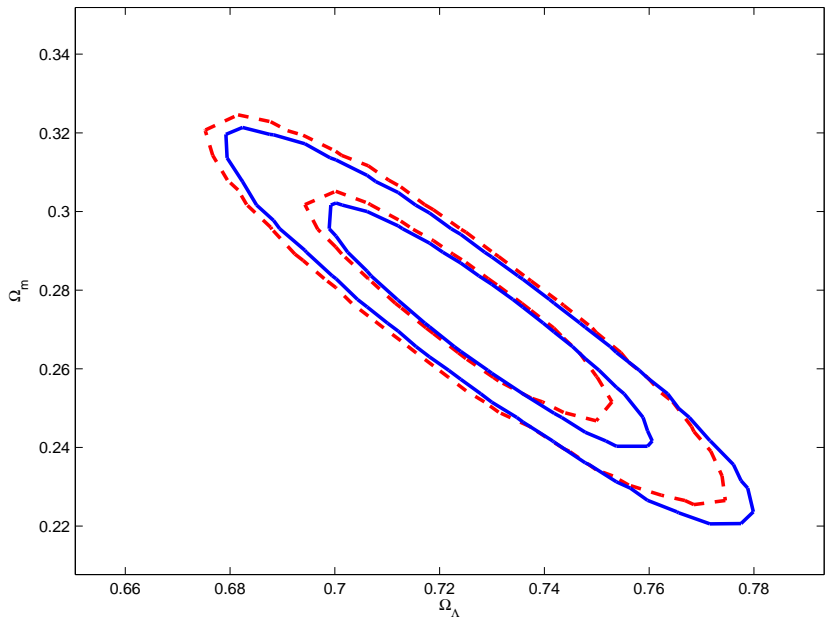}
\caption{Color Online: $2D$ contours plots in the
$\Omega_{\Lambda}-\Omega_m$ plane. The solid blue lines denote the
$1\sigma$ and $2\sigma$ regions from SN Union 2 alone. The red
dashed lines denote the $1\sigma$ and $2\sigma$ regions from the
combination of SN Union 2 and $109$ GRBs.}\label{fig:con2dsngrb}
\end{figure}

\begin{figure}[tbh]
\centering
\includegraphics[width=5.0in]{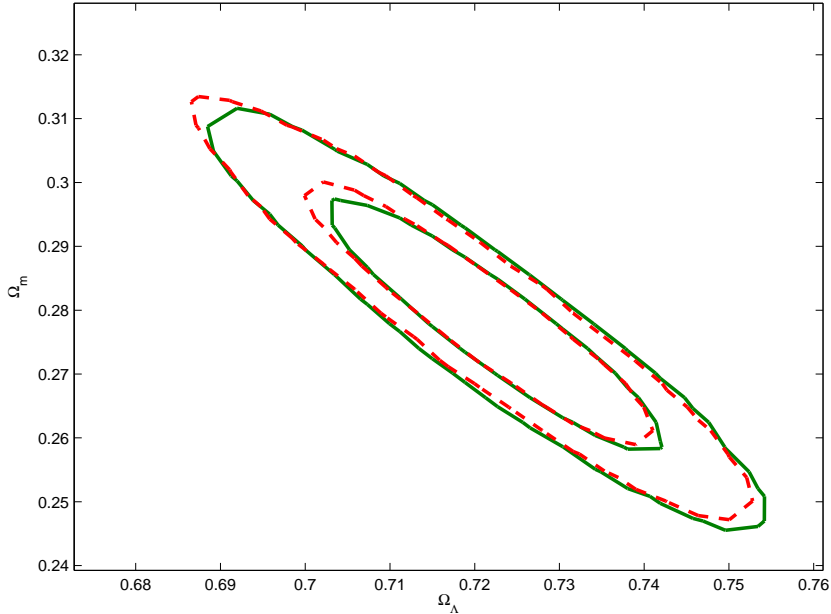}
\caption{Color
Online: $2D$ contours plots in the $\Omega_{\Lambda}-\Omega_m$
plane. The solid blue lines denote the $1\sigma$ and $2\sigma$
regions from SN+BAO+CMB. The red dashed lines denote the
$1\sigma$ and $2\sigma$ regions from the combination of SN+BAO+CMB+GRBs.}\label{fig:conall}
\end{figure}

\begin{figure}[tbh]
\centering
\includegraphics[width=5.0in]{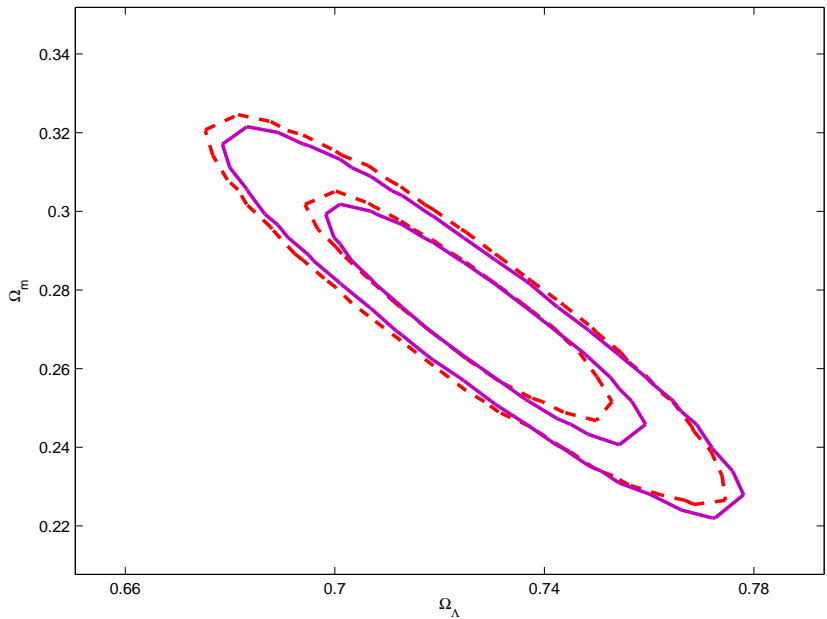}
\caption{Color
Online: $2D$ contours plots in the $\Omega_{\Lambda}-\Omega_m$
plane. The solid pink lines denote the $1\sigma$ and $2\sigma$
regions from SN Union 2 + $59$ GRBs from Wei's results. The red dashed lines denote the
$1\sigma$ and $2\sigma$ regions from the combination of SN+GRBs in our case.}\label{fig:concompa}
\end{figure}

In our calculations, we have taken the total likelihood function
$L\propto e^{-\chi^2/2}$ to be the products of the separate
likelihoods of SN, BAO, CMB and GRBs. Then we get $\chi^2$ 
\begin{eqnarray}
\chi^2=\chi^2_{SN}+\chi^2_{BAO}+\chi^2_{CMB}+\chi^2_{GRB},
\end{eqnarray}
where the separate likelihoods of SN, BAO, CMB are shown in the Appendix \ref{sec:appendix}. The
$\chi^2_{GRB}$ is the form of Eq. (\ref{eq:chi2GRB}). In this work, we only consider the $\Lambda$CDM model as a simple example. Its generalization to constrain other cosmological models is straight forward.
In the case, when SN Ia and
GRBs are combined as cosmic constraint, we have $\chi^2_{min}=547.727$ and
the best fit values of model parameter
$\Omega_{m0}=0.274^{+0.0203}_{-0.0194}$. If the SN Ia is used alone,
one has $\chi^2_{min}=542.680$ and
$\Omega_{m0}=0.270^{+0.0222}_{-0.0213}$. The corresponding contour
plot is shown in Fig. \ref{fig:con2dsngrb}.
When combing the cosmic observations considered in this work
together, we have the resulted $\chi^2_{min}=549.383$ and $\Omega_{m0}=0.279^{+0.0138}_{-0.0131}$.
Also, the best fit values of $\Omega_{m0}=0.277^{+0.0132}_{-0.0128}$ and $\chi^2_{min}=544.451$ when GRBs is not used for comparison. The $2D$ contour plots are shown in Fig. \ref{fig:conall}.
We can find that when GRBs is used, the relative errors of model parameters is shrunken. 
Here, we can compare our result with Wei's one where he used $50$ low redshift GRBs to calibrate the Amati relation. So, we add $59$ high redshift data points of GRBs to SN Union 2 data sets.
After  constraint via MCMC, we have $\chi^2_{min}=565.919$ and $\Omega_{m0}=0.271^{+0.0198}_{-0.0192}$. The corresponding $2D$ contour plot is shown in Fig. \ref{fig:concompa}. The constrained results show that relative large values of $\Omega_{m0}$ are favored when GRBs data points are employed. This is consistent with the clues shown in Fig. \ref{fig:errorbars}.

\section{Summary and Discussion}

In this paper, by using $109$ GRBs data points via Amati relation, we have derived five data points of distance measurements which {\it do not} depend on any cosmological models, i.e. in a
model-independent way, based on the method firstly advocated by Wang \cite{ref:wang}. Then it can be used to constrain any other cosmological models without the so-called circular and data sets consistence problem. We also find out that the concordance model ($\Omega_{m0}=0.27$) is almost at the boundary of $1\sigma$ regions of data points. So, one can reject the null hypothesis that Amati relation is based on cosmology slightly greater than $4.5\sigma$ as shown in Fig. \ref{fig:errorbars}.

When GRBs (the five data points of distance measurements) is used as a complementary cosmic constraint to $\Lambda$CDM model as an example, the errors of model parameters are shrunken. Though the constraint is not much tighter than that calibrated via low redshift SN, the difference is very small. And, the results can be used to constrain other cosmological model beyond $\Lambda$CDM. The important thing is that it {\it does not} depend on SN Ia data set. The merits of this method has been mentioned in the introduction. Here, we list possible potential drawbacks of this 
completely model in-dependent method as follows: (i). The
constraint is not much tighter than that obtained from calibration
by using SN Ia data points. The looser may come from the lack of
enough GRBs data points or the number of bins is not larger. (ii).
How to design a cosmological model indicator, here it is model independent distance measurement $\bar{r}_p(z)$, which is sensitive to
distinguish the cosmological models may be the main points of this
method. In this way, the lack of having no enough GRBs data
points can be avoided in some senses. (iii) Another negative effect comes from the choice of fiducial  redshift $z_0$. Some average values of $r_p(z_0)$ would be better. 

At last we have to alert the readers that our analysis is based on the assumption that Amati relation is reliable. However, in the literatures, the Amati relation has been criticized for many reasons as mentioned in the introduction. If that is true, our analysis would be unreliable.

\acknowledgments The author thanks an anonymous referee for invaluable improvement of this paper and Dr. Jirong Mao for useful discussion on Amati relation. This work is supported by the Fundamental Research Funds for the Central Universities (DUT10LK31) and (DUT11LK39) of P.R. China.

\appendix

\section{Cosmic Observations: SN, BAO and CMB}\label{sec:appendix}

\subsection{Type Ia Supernovae constraints}

Recently, SCP (Supernova Cosmology Project) collaboration released
their Union2 dataset which consists of 557 SN Ia \cite{ref:SN557}.
The distance modulus $\mu(z)$ is defined as
\begin{equation}
\mu_{th}(z)=5\log_{10}[\bar{d}_{L}(z)]+\mu_{0},
\end{equation}
where $\bar{d}_L(z)$ is the Hubble-free luminosity distance $H_0
d_L(z)/c=H_0 d_A(z)(1+z)^2/c$, with $H_0$ the Hubble constant,
defined through the re-normalized quantity $h$ as $H_0=100 h~{\rm km
~s}^{-1} {\rm Mpc}^{-1}$, and $\mu_0\equiv42.38-5\log_{10}h$.
Additionally, the observed distance moduli $\mu_{obs}(z_i)$ of SN Ia
at $z_i$ is
\begin{equation}
\mu_{obs}(z_i) = m_{obs}(z_i)-M,
\end{equation}
where $M$ is their absolute magnitudes.

For the SN Ia dataset, the best fit values of the parameters $p_s$
can be determined by a likelihood analysis, based on the calculation
of
\begin{eqnarray}
\chi^2(p_s,M^{\prime})&\equiv& \sum_{SN}\frac{\left\{
\mu_{obs}(z_i)-\mu_{th}(p_s,z_i)\right\}^2} {\sigma_i^2}  \nonumber\\
&=&\sum_{SN}\frac{\left\{ 5 \log_{10}[\bar{d}_L(p_s,z_i)] -
m_{obs}(z_i) + M^{\prime} \right\}^2} {\sigma_i^2}, \label{eq:chi2}
\end{eqnarray}
where $M^{\prime}\equiv\mu_0+M$ is a nuisance parameter which
includes the absolute magnitude and the parameter $h$. The nuisance
  parameter $M^{\prime}$ can be marginalized over
analytically \cite{ref:SNchi2} as
\begin{equation}
\bar{\chi}^2(p_s) = -2 \ln \int_{-\infty}^{+\infty}\exp \left[
-\frac{1}{2} \chi^2(p_s,M^{\prime}) \right] dM^{\prime},\nonumber
\label{eq:chi2marg}
\end{equation}
resulting to
\begin{equation}
\bar{\chi}^2 =  A - \frac{B^2}{C} + \ln \left( \frac{C}{2\pi}\right)
, \label{eq:chi2mar}
\end{equation}
with
\begin{eqnarray}
&&A=\sum_{SN} \frac {\left\{5\log_{10}
[\bar{d}_L(p_s,z_i)]-m_{obs}(z_i)\right\}^2}{\sigma_i^2},\nonumber\\
&& B=\sum_{SN} \frac {5
\log_{10}[\bar{d}_L(p_s,z_i)]-m_{obs}(z_i)}{\sigma_i^2},\nonumber
\\
&& C=\sum_{SN} \frac {1}{\sigma_i^2}\nonumber.
\end{eqnarray}
Relation (\ref{eq:chi2}) has a minimum at the nuisance parameter
value $M^{\prime}=B/C$, which contains information of the values of
$h$ and $M$. Therefore, one can extract the values of $h$ and $M$
provided the knowledge of one of them. Finally, note that the
expression
\begin{equation}
\chi^2_{SN}(p_s,B/C)=A-(B^2/C),\label{eq:chi2SN}\nonumber
\end{equation}
which coincides to Eq. (\ref{eq:chi2mar}) up to a constant, is often
used in the likelihood analysis \cite{ref:smallomega,ref:SNchi2},
and thus in this case the results will not be affected by a flat
$M^{\prime}$ distribution.

\subsection{Baryon Acoustic Oscillation constraints}

The Baryon Acoustic Oscillations are detected in the clustering of
the combined the 2dF Galaxy Redshift Survey (2dFGRS) and Sloan Digital Sky Survey (SDSS) main galaxy samples, and measure the
distance-redshift relation at $z = 0.2$. Additionally, Baryon
Acoustic Oscillations in the clustering of the SDSS luminous red
galaxies measure the distance-redshift relation at $z = 0.35$. The
observed scale of the BAO calculated from these samples, as well as
from the combined sample, are jointly analyzed using estimates of
the correlated errors to constrain the form of the distance measurement
$D_V(z)$ \cite{ref:Okumura2007,ref:Percival2,ref:Eisenstein05}
\begin{equation}
D_V(z)=c\left(\frac{z}{\Omega_k
H(z)}\mathrm{sinn}^2[\sqrt{|\Omega_k|}\int_0^z\frac{dz'}{H(z')}]\right)^{1/3}.
\label{eq:DV}
\end{equation}
where $\mathrm{sinn}(x)=\sin(x), x,\sinh(x)$ for $\Omega_k<0$, $\Omega_k=0$ and $\Omega_k>0$ respectively. The peak positions of the BAO depend on the ratio of $D_V(z)$ to the
sound horizon size at the drag epoch (where baryons were released
from photons) $z_d$, which can be obtained by using a fitting
formula \cite{ref:Eisenstein}:
\begin{eqnarray}
&&z_d=\frac{1291(\Omega_mh^2)^{-0.419}}{1+0.659(\Omega_mh^2)^{0.828}}[1+b_1(\Omega_bh^2)^{b_2}],
\end{eqnarray}
with
\begin{eqnarray}
&&b_1=0.313(\Omega_mh^2)^{-0.419}[1+0.607(\Omega_mh^2)^{0.674}], \\
&&b_2=0.238(\Omega_mh^2)^{0.223}.
\end{eqnarray}
In this paper, we use the data of $r_s(z_d)/D_V(z)$ extracted from
the Sloan Digitial Sky Survey (SDSS) and the Two Degree Field Galaxy
Redshift Survey (2dFGRS) \cite{ref:Percival3}, which are listed in
Table \ref{baodata}, where $r_s(z)$ is the comoving sound horizon
size
\begin{eqnarray}
r_s(z)&&{=}c\int_0^t\frac{c_sdt}{a}=c\int_0^a\frac{c_sda}{a^2H}=c\int_z^\infty
dz\frac{c_s}{H(z)} \nonumber\\
&&{=}\frac{c}{\sqrt{3}}\int_0^{1/(1+z)}\frac{da}{a^2H(a)\sqrt{1+(3\Omega_b/(4\Omega_\gamma)a)}},
\end{eqnarray}
where $c_s$ is the sound speed of the photon$-$baryon fluid
\cite{ref:Hu1, ref:Hu2, ref:Caldwell}:
\begin{eqnarray}
&&c_s^{-2}=3+\frac{4}{3}\times\frac{\rho_b(z)}{\rho_\gamma(z)}=3+\frac{4}{3}\times(\frac{\Omega_b}{\Omega_\gamma})a,
\end{eqnarray}
and here $\Omega_\gamma=2.469\times10^{-5}h^{-2}$ for
$T_{CMB}=2.75K$.

\begin{table}[htbp]
\begin{center}
\begin{tabular}{c|l}
\hline\hline
 $z$ &\ $r_s(z_d)/D_V(z)$  \\ \hline
 $0.2$ &\ $0.1905\pm0.0061$  \\ \hline
 $0.35$  &\ $0.1097\pm0.0036$  \\
\hline
\end{tabular}
\end{center}
\caption{\label{baodata} The observational $r_s(z_d)/D_V(z)$
data~\cite{ref:Percival2}.}
\end{table}
Using the data of BAO in Table \ref{baodata} and the inverse
covariance matrix $V^{-1}$ in \cite{ref:Percival2}:

\begin{eqnarray}
&&V^{-1}= \left(
\begin{array}{cc}
 30124.1 & -17226.9 \\
 -17226.9 & 86976.6
\end{array}
\right),
\end{eqnarray}

Thus, the $\chi^2_{BAO}(p_s)$ is given as
\begin{equation}
\chi^2_{BAO}(p_s)=X^tV^{-1}X,\label{eq:chi2BAO}
\end{equation}
where $X$ is a column vector formed from the values of theory minus
the corresponding observational data, with
\begin{eqnarray}
&&X= \left(
\begin{array}{c}
 \frac{r_s(z_d)}{D_V(0.2)}-0.190533 \\
 \frac{r_s(z_d)}{D_V(0.35)}-0.109715
\end{array}
\right),
\end{eqnarray}
and $X^t$ denotes its transpose.

\subsection{Cosmic Microwave Background constraints}

The CMB shift parameter $R$ is provided by \cite{ref:Bond1997}
\begin{equation}
R(z_{\ast})=\frac{\sqrt{\Omega_m
H^2_0}}{\sqrt{|\Omega_k|}}\mathrm{sinn}[\sqrt{|\Omega_k|}\int_0^{z{_\ast}}\frac{dz'}{H(z')}],
\end{equation}
here, the redshift $z_{\ast}$ (the decoupling epoch of photons) is
obtained using the fitting function \cite{Hu:1995uz}
\begin{equation}
z_{\ast}=1048\left[1+0.00124(\Omega_bh^2)^{-0.738}\right]\left[1+g_1(\Omega_m
h^2)^{g_2}\right],\nonumber
\end{equation}
where the functions $g_1$ and $g_2$ read
\begin{eqnarray}
g_1&=&0.0783(\Omega_bh^2)^{-0.238}\left(1+ 39.5(\Omega_bh^2)^{0.763}\right)^{-1},\nonumber \\
g_2&=&0.560\left(1+ 21.1(\Omega_bh^2)^{1.81}\right)^{-1}.\nonumber
\end{eqnarray}
In additional, the acoustic scale is related to the first distance
ratio and is expressed as
\begin{eqnarray}
&&l_A=\frac{\pi}{r_s(z_{\ast})}\frac{c}{\sqrt{|\Omega_k|}}\mathrm{sinn}[\sqrt{|\Omega_k|}\int_0^{z_\ast}\frac{dz'}{H(z')}].
\end{eqnarray}

\begin{table}[htbp]
\begin{center}
\begin{tabular}{c|ccc}
\hline\hline
  &\ $\mathrm{7-year}$ $\mathrm{ML}$ &\ $\mathrm{7-year}$ $\mathrm{mean}$ &\ $\mathrm{error}$, $\mathrm{\sigma}$ \\ \hline
 $l_A(z_\ast)$ &\ $302.09$ &\ $302.69$ &\ $0.76$ \\ \hline
 $R(z_\ast)$ &\ $1.725$ &\ $1.726$ &\ $0.018$ \\ \hline
 $z_{\ast}$  &\ $1091.3$ &\ $1091.36$ &\ $0.91$ \\
\hline
\end{tabular}
\end{center}
\caption{\label{cmbdata} The observational $l_A, R, z_{\ast}$
data~\cite{CMB:7yr}. The $\mathrm{ML}$ values are used in this work
as recommended.}
\end{table}

Using the data of $l_A, R, z_{\ast}$ in \cite{CMB:7yr}, which are
listed in Table \ref{cmbdata}, and their covariance matrix of
$[l_A(z_\ast), R(z_\ast), z_{\ast}]$ referring to \cite{CMB:7yr}:
\begin{eqnarray}
&&C^{-1}= \left(
\begin{array}{ccc}
2.305 & 29.698 & -1.333\\
 29.698 & 6825.270 & -113.180\\
 -1.333 & -113.180 & 3.414
\end{array}
\right),
\end{eqnarray}
we can calculate the likelihood $L$ as $\chi^2_{CMB}=-2\ln L$:
\begin{eqnarray}
&&\chi^2_{CMB}=\bigtriangleup d_i[C^{-1}(d_i,d_j)][\bigtriangleup
d_i]^t,
\end{eqnarray}
where $\bigtriangleup d_i=d_i-d_i^{data}$ is a row vector, and
$d_i=(l_A, R, z_{\ast})$.


\begin{thebibliography}{*}

\bibitem{ref:Riess98} A. G. Riess, {\it et al.}, {\it Astron. J.}
{\bf 116} 1009(1998) [astro-ph/9805201].

\bibitem{ref:Perlmuter99} S. Perlmutter, {\it et al.}, {\it Astrophys. J.} {\bf
517} 565(1999) [astro-ph/9812133].

\bibitem{ref:Schaefer} B. E. Schaefer, Astrophys. J. 660, 16 (2007) [astro-ph/0612285].

\bibitem{ref:Lihong} H. Li, J. Q. Xia, J. Liu, G. B. Zhao, Z. H. Fan, X. Zhang, Astrophys. J. 680, 92(2008).

\bibitem{ref:cosmography}  S. Capozziello, L. Izzo, Astron. Astrophys, 490, 31( 2008); V. Vitagliano, J. Q. Xia, S. Liberati, M. Viel, JCAP03(2010)005.

\bibitem{ref:Liang} N. Liang, W. K. Xiao, Y. Liu and S. N. Zhang, Astrophys. J. 685, 354 (2008)
[arXiv:0802.4262]; N. Liang and S. N. Zhang, AIP Conf. Proc. 1065,
367 (2008) [arXiv:0808.2655]; T. S. Wang and N. Liang,
arXiv:0910.5835 [astro-ph.CO]; N. Liang, P. Wu and S. N. Zhang,
Phys. Rev. D 81, 083518 (2010) [arXiv:0911.5644]; H. Gao, N. Liang
and Z. H. Zhu, arXiv:1003.5755 [astro-ph.CO].

\bibitem{ref:Wei} H. Wei and S. N. Zhang, Eur. Phys. J. C 63, 139 (2009) [arXiv:0808.2240];

\bibitem{ref:Wei109} H. Wei, JCAP1008:020(2010), arXiv:1004.4951 [astro-ph.CO].

\bibitem{ref:Amati'srelation} L. Amati et al., Astron. Astrophys. 390, 81 (2002) [astro-ph/0205230].

\bibitem{ref:errors}  Herman J. Mosquera Cuesta, Habib Dumet M., Cristina
Furlanetto, JCAP0807,004(2008).

\bibitem{r16}
L.~Amati {\it et al.},
 Mon.\ Not.\ Roy.\ Astron.\ Soc.\  {\bf 391}, 577 (2008)
 [arXiv:0805.0377].

\bibitem{r17}
L.~Amati, arXiv:1002.2232 [astro-ph.HE]; L. Amati, Mon. Not. Roy.
Astron. Soc. 372, 233 (2006) [astro-ph/0601553].

\bibitem{r18}
L.~Amati, F.~Frontera and C.~Guidorzi, arXiv:0907.0384
[astro-ph.HE].


\bibitem{ref:Li} L.-X. Li, Mon.Not.Roy.Astron.Soc., Lett, 374, L20(2006).

\bibitem{ref:Nakar}  E. Nakar, T. Piran, Mon.Not.Roy.Astron.Soc., Lett, 360, L73(2005).

\bibitem{ref:Band}  D. L. Band, R.D. Preece, ApJ, 627, 319(2005). 

\bibitem{ref:Collazzi} A. C. Collazzi, B. E. Schaefer, A. Goldstein, R.D. Preece, 	arXiv:1112.4347[astro-ph.HE].

\bibitem{ref:wang} Y. Wang, Phys.Rev.D 78,123532(2008).

\bibitem{ref:MCMC} http://cosmologist.info/cosmomc/; A. Lewis and S. Bridle, Phys. Rev. D 66, 103511 (2002).

\bibitem{essence} W.~M.~Wood-Vasey {\it et al.}, Astrophys. J. {\bf 666} 694 (2007) [astro-ph/0701041].


\bibitem{ref:SN557} R. Amanullah et al. [Supernova Cosmology Project Collaboration], arXiv:1004.1711
[astro-ph.CO].

\bibitem{ref:SNchi2}
S. Nesseris and L. Perivolaropoulos, Phys. Rev. D {\bf  72} 123519
(2005); L. Perivolaropoulos, Phys. Rev. D {\bf 71} 063503 (2005); E.
Di Pietro and J. F. Claeskens, Mon. Not. Roy. Astron. Soc. {\bf 341}
1299 (2003); A.~C.~C.~Guimaraes, J.~V.~Cunha and J.~A.~S.~Lima, JCAP
{\bf 0910} 010 (2009).

\bibitem{ref:smallomega} E.~Garcia-Berro, E.~Gaztanaga, J.~Isern, O.~Benvenuto and
L.~Althaus, astro-ph/9907440; A. Riazuelo and J. Uzan, Phys. Rev. D
{\bf 66} 023525 (2002); V. Acquaviva and L. Verde, JCAP {\bf 0712}
001 (2007).


\bibitem{ref:Okumura2007}
T.~Okumura, T.~Matsubara, D.~J.~Eisenstein, I.~Kayo, C.~Hikage,
A.~S.~Szalay and D.~P.~Schneider, Astrophys. J.  {\bf 676} 889
(2008).

\bibitem{ref:Percival2} W. J. Percival {\it et al.}, arXiv:0907.1660 [astro-ph.CO].

\bibitem{ref:Eisenstein05} D. J. Eisenstein {\it et al.}, [SDSS Collabaration], Astrophys. J. {\bf 633} 560 (2005)
[astro-ph/0501171].

\bibitem{ref:Eisenstein} D. J. Eisenstein and W. Hu, Astrophys. J. {\bf496} 605 (1998).

\bibitem{ref:Percival3} W. J. Percival {\it et al.}, Mon. Not. R. Astron. Soc. {\bf 381} 1053 (2007) arXiv:0705.3323 [astro-ph.CO].

\bibitem{ref:Hu1} W. Hu and N. Sugiyama, Astrophys. J. {\bf 444} 489 (1995) [arXiv:astro-ph/9407093].

\bibitem{ref:Hu2} W. Hu, M. Fukugita, M. Zaldarriaga and M. Tegmark, Astrophys. J. {\bf 549} 669 (2001) [arXiv:astro-ph/0006436].

\bibitem{ref:Caldwell} R. R. Caldwell and M. Doran, Phys. Rev. D {\bf 69} 103517 (2004).


\bibitem{ref:Bond1997} J.~R.~Bond, G.~Efstathiou and M.~Tegmark, Mon. Not. Roy. Astron. Soc.  {\bf 291} L33 (1997).

\bibitem{Hu:1995uz} W.~Hu and N.~Sugiyama, Astrophys. J. {\bf 471} 542 (1996).

\bibitem{CMB:7yr} E. Komatsu et al. [WMAP Collaboration], arXiv:1001.4538 [astro-ph.CO].

\end{thebibliography}
\end{document}